\documentclass[twocolumn,showpacs,preprintnumbers,amsmath,amssymb,superscriptaddress]{revtex4}

\usepackage{graphicx}
\usepackage{dcolumn}
\usepackage{bm}

\newcommand{\nn}{\nonumber}
\newcommand{\ovl}[1]{\overline{#1}}

\newcommand{\p}{\partial}
\newcommand{\pslash}{p\kern-1ex /}
\newcommand{\lslash}{l\kern-1ex /}
\newcommand{\kslash}{k\kern-1ex /}
\newcommand{\dslash}{\p\kern-1.2ex /}
\newcommand{\Dslash}{{\cal D}\kern-1.5ex /}
\newcommand{\Aslash}{A\kern-1.2ex 23/}

\newcommand{\hu}[1]{{\bf }}

\newcommand{\les}{\stackrel{<}{{}_{\sim}}}

\newcommand{\YITP}{
  Yukawa Institute for Theoretical Physics,
  Kyoto University, Kyoto 606-8502, Japan
}

\newcommand{\RBRC}{
  Riken BNL Research Center, 
  Upton, NY 11973, USA
}

\newcommand{\GUAS}{
  School of High Energy Accelerator Science,
  The Graduate University for Advanced Studies (Sokendai),
  Tsukuba 305-0801, Japan
}

\newcommand{\KEK}{
  High Energy Accelerator Research Organization (KEK),
  Tsukuba 305-0801, Japan
}

\newcommand{\TSUKUBA}{
  Graduate School of Pure and Applied Sciences,
  University of Tsukuba, Tsukuba 305-8571, Japan
}

\newcommand{\NTU}{
  Physics Department, Center for Theoretical Sciences, 
  and National Center for Theoretical Sciences, National Taiwan
  University,
  Taipei 10617, Taiwan
}

\newcommand{\NBI}{
 The Niels Bohr Institute,
 The Niels Bohr International Academy,
 Blegdamsvej 17 DK-2100 Copenhagen {\O}
 Denmark
}

\newcommand{\RCAS}{
  Research Center for Applied Sciences, Academia Sinica,
  Taipei~115, Taiwan
}

\begin{document}

\preprint{KEK-CP-212,NTUTH-08-505A,UT-559,YITP-08-38}

\title{
  Convergence of the chiral expansion in two-flavor lattice QCD
}

\author{J.~Noaki}
\affiliation{\KEK}
\author{S.~Aoki}
\affiliation{\TSUKUBA}
\affiliation{\RBRC}
\author{T.W.~Chiu}
\affiliation{\NTU}
\author{H.~Fukaya}
\affiliation{\NBI}
\affiliation{\KEK}
\author{S.~Hashimoto}
\affiliation{\KEK}
\affiliation{\GUAS}
\author{T.H.~Hsieh}
\affiliation{\RCAS}
\author{T.~Kaneko}
\affiliation{\KEK}
\affiliation{\GUAS}
\author{H.~Matsufuru}
\affiliation{\KEK}
\author{T.~Onogi}
\affiliation{\YITP}
\author{E.~Shintani}
\affiliation{\KEK}
\author{N.~Yamada}
\affiliation{\KEK}
\affiliation{\GUAS}

\collaboration{JLQCD and TWQCD Collaborations}

\date{\today}

\begin{abstract}
  We test the convergence property of the chiral perturbation theory
  (ChPT) using a lattice QCD calculation of pion mass and decay
  constant with two dynamical quark flavors.
  The lattice calculation is performed using the overlap fermion
  formulation, which realizes exact chiral symmetry at finite lattice
  spacing. 
  By comparing various expansion prescriptions, we find that the
  chiral expansion is well saturated at the next-to-leading order (NLO) 
  for pions lighter than $\sim$450~MeV.
  Better convergence behavior is found in particular for a resummed
  expansion parameter $\xi$, with which the lattice data in the pion
  mass region 290$\sim$750~MeV can be fitted well with the
  next-to-next-to-leading order (NNLO) formulae.
  We obtain the results in two-flavor QCD for the low energy constants
  $\bar{l}_3$ and $\bar{l}_4$ as well as the pion decay constant, the
  chiral condensate, and the average up and down quark mass.
\end{abstract}


\maketitle

Chiral perturbation theory (ChPT) is a powerful tool to analyze the
dynamics of low energy pions \cite{Gasser:1983yg}.
The expansion parameter in ChPT is the pion mass (or momentum) 
divided by the typical scale of the underlying theory such as
Quantum Chromodynamics (QCD).
Good convergence of the chiral expansion is observed for physical
pions in the analysis including the
next-to-next-to-leading order (NNLO) for the pion-pion scattering
\cite{Colangelo:2001df}, for instance.
In the kaon mass region, on the other hand, the validity of ChPT is
not obvious and in fact an important issue in many phenomenological 
applications.

Lattice QCD calculation can, in principle, be used for a detailed test 
of the convergence property of ChPT, as one can freely vary the quark
mass, typically in the range $m_s/5\sim m_s$ with $m_s$ the physical
strange quark mass.
However, such a direct test has been difficult, since the lattice
regularization of the quark action explicitly violates flavor and/or
chiral symmetry in the conventional formulations, such as the
Wilson and staggered fermions.
One then has to introduce additional terms with unknown parameters in
order to describe those violations in ChPT, hence the test requires
precise continuum extrapolation.

The aim of this article is to provide a direct comparison between
the ChPT predictions and lattice QCD calculations, using the overlap
fermion formulation on the lattice, that preserves exact chiral
symmetry at finite lattice spacing $a$
~\cite{Neuberger:1997fp,Neuberger:1998wv}.
With the exact chiral symmetry, the use of the continuum ChPT is
valid to describe the lattice data at a finite lattice spacing up
to Lorentz violating corrections; the discretization error of
${\cal O}(a^2)$ affects the value of the Low Energy Constants
(LECs) and unknown Lorentz violating corrections.
In order to make a cleaner analysis, we consider two-flavor QCD in
this work, leaving the similar study in 2+1-flavor QCD, which
introduces much more complications, for a future work.
We calculate the pion mass and decay constant, for
which the NNLO calculations are available in ChPT
\cite{Bijnens:1998fm}.
A preliminary report of this work is found in \cite{Noaki:2007es}.

Lattice simulations are performed on a 
$L_s^3\times L_t = 16^3\times 32$ lattice at a lattice spacing
$a$ = 0.1184(03)(21)~fm determined with an input $r_0$ = 0.49~fm,
the Sommer scale defined for the heavy quark potential.
At six different sea quark masses $m_{\rm sea}$, covering the pion
mass region
$290\ {\rm MeV} \les m_\pi \les 750\ {\rm MeV}$,
we generate 10,000 trajectories, among which the calculation of the
pion correlator is carried out at every 20 trajectories.
For further details of the simulation we refer \cite{Aoki:2008tq}. 

In the calculation of the pion correlator, we computed in advance the
lowest 50 conjugate pairs of eigenmode of the overlap-Dirac operator
on each gauge configuration and stored on the disk.
Then, by using the eigenmodes to construct a preconditioner, the
inversion of the overlap-Dirac operator can be done with 
only $\approx$ 15\% of the CPU time of the full calculation.
The low-modes are also used to improve the statistical accuracy by
averaging their contribution to the correlators over 32 source points
distributed in each time slice.
The correlators are calculated with a point source and a smeared
source; the pion mass $m_\pi$ and decay constant $f_\pi$ are obtained
from a simultaneous fit of them.

The pion decay constant $f_\pi$ is defined through
$\langle 0|{\cal A}_\mu|\pi(p)\rangle = if_\pi p_\mu$, where ${\cal A}_\mu$
is the (continuum) iso-triplet axial-vector current.
Instead of ${\cal A}_\mu$, we calculate the matrix element of 
pseudo-scalar density $P^{\rm lat}$ on the lattice
using the PCAC relation $\partial_\mu {\cal A}_\mu = 2m_q^{\rm lat}
P^{\rm lat}$ with $m_q^{\rm lat}$ the bare quark mass.
Since the combination $m_q^{\rm lat}P^{\rm lat}$ is not renormalized,
no renormalization factor is needed in the calculation of $f_\pi$.
This is possible only when the chiral symmetry is exact.
The renormalization factor for the quark mass 
$m_q = Z_m^{\ovl{\rm MS}}(\mathrm{2~GeV})\, m_q^{\rm lat}$ is calculated
non-perturbatively through the RI/MOM scheme, with which the
renormalization condition is applied at some off-shell momentum for
propagators and vertex functions.
Such a non-perturbative calculation suffers from the non-trivial quark 
mass dependence of the chiral condensate.
By using the calculated low-modes explicitly, we are able to control the 
mass dependence to determine $Z_m^{\rm RI/MOM}$ more reliably. 
In the chiral limit, we obtain 
$Z_m^{\ovl{\rm MS}}(\mathrm{2~GeV})= 0.838(14)(03)$, 
where the second error arises from a subtraction of power divergence 
from the chiral condensate.
The details of this calculation will be given elsewhere.

Since our numerical simulation is done on a finite volume lattice 
with $m_\pi L_s \simeq 2.9$ for the lightest sea quark, the finite volume
effect could be significant. 
We make a correction for the finite volume effect using the estimate within ChPT
calculated up to ${\cal O}(m_\pi^4/(4\pi f_\pi)^4)$ \cite{Colangelo:2005gd}.
The size of the corrections for $m_\pi^2$ and $f_\pi$ is about 5\% for 
the lightest pion mass and exponentially suppressed for heavier data points.
In addition, there is a correction due to fixing the global
topological charge in our simulation \cite{Aoki:2008tq, Fukaya:2006vs}.
This leads to a finite volume effect of ${\cal O}(1/V)$ with $V$ the physical
space-time volume.
The correction is calculable within ChPT
\cite{Brower:2003yx,Aoki:2007ka} depending on the value of topological
susceptibility $\chi_t$, which we calculated in \cite{Aoki:2007pw}.
At NLO, the correction for $m_\pi^2$ is similar in magnitude but
opposite in sign to the ordinary finite volume effect at the
lightest pion mass, and thus almost cancels.
For $f_\pi$ the finite volume effect due to the fixed topology starts
at NLO and therefore is a subdominant effect.
Note that the LECs appear in the calculation of these correction
factors. 
We use their phenomenological values at the scale of physical (charged) 
pion mass $m_{\pi^+}=139.6$ MeV:
$\bar{l}_1^{\rm phys}=-0.4\pm 0.6$, 
$\bar{l}_2^{\rm phys}= 4.3\pm 0.1$, 
$\bar{l}_4^{\rm phys}= 4.4\pm 0.2$, determined at the NNLO
~\cite{Colangelo:2001df} and $\bar{l}_3^{\rm phys}= 2.9\pm 2.4$.
The errors in these values are reflected in the following analysis
assuming a gaussian distribution.

\begin{figure}
  \begin{center}
   \includegraphics[width=7.2cm,clip]{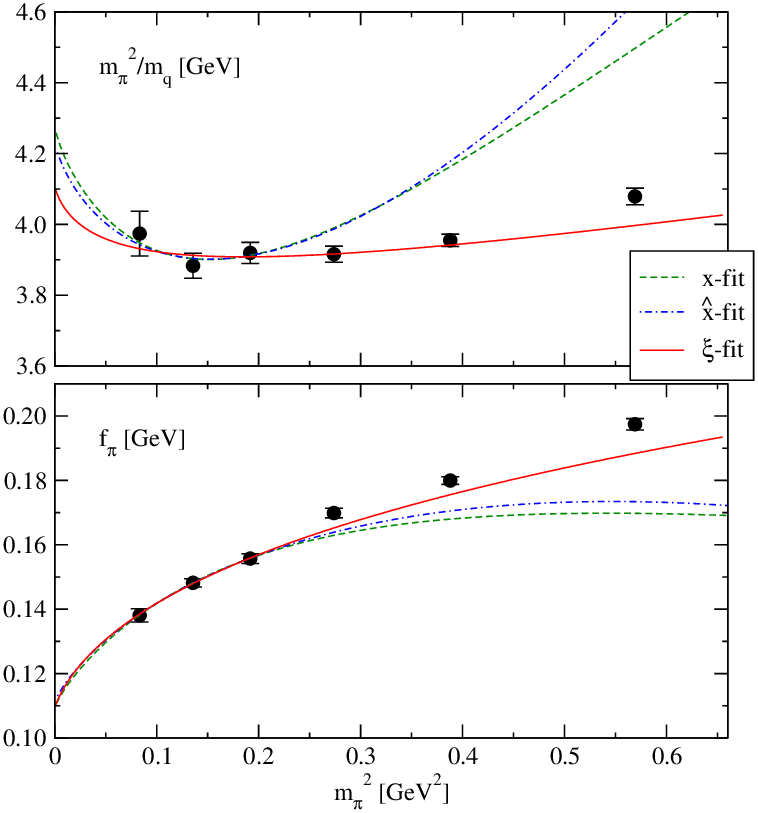}
  \end{center}
 \vspace{-0.5cm}
 \caption{
 Comparison of the chiral fits including the NLO terms
   for $m_\pi^2/m_q$ (top) and $f_\pi$ (bottom). 
   Fit curves to three lightest data points obtained with different
   choices of the expansion parameter ($x$, $\hat{x}$, and $\xi$) are
   shown as a function of $m_\pi^2$.
 } \label{FIT3pts}
\end{figure}

After applying the finite volume corrections, we first analyze the 
numerical data for $m_\pi^2/m_q$ and $f_\pi$ using the ChPT formulae at NLO,
\begin{eqnarray}
  m_\pi^2/m_q &=& 2B(1+\tfrac{1}{2}x\ln x)+c_3 x,
  \label{M_NLO}\\
  f_\pi &=& f(1-x\ln x)+c_4 x,
  \label{F_NLO}
\end{eqnarray}
where $f$ is the pion decay constant in the chiral limit and $B$ is
related to the chiral condensate.
Here the expansion is made in terms of 
$x\equiv 4 B m_q/(4\pi f)^2$.
The parameters $c_3$ and $c_4$ are related to the LECs 
$\bar{l}_3^{\rm phys}$ and $\bar{l}_4^{\rm phys}$, respectively.
At NLO, {\it i.e.} ${\cal O}(x)$, these expressions are unchanged when 
one replaces the expansion parameter $x$ by
$\hat{x} \equiv 2m_\pi^2/(4\pi f)^2$ or 
$\xi \equiv 2m_\pi^2/(4\pi f_\pi)^2$,
where $m_\pi$ and $f_\pi$ denote those at a finite quark mass.
Therefore, in a small enough pion mass region the three 
expansion parameters should describe the lattice data equally well.

Three fit curves ($x$-fit, $\hat{x}$-fit, and $\xi$-fit) for the three
lightest pion mass points ($m_\pi\lesssim$ 450~MeV) are shown in
Figure~\ref{FIT3pts} as a function of $m_\pi^2$.
From the plot we observe that the different expansion parameters
seem to describe the three lightest points equally well; 
the values of $\chi^2/$dof are 0.30, 0.33 and 0.66 for $x$-,
$\hat{x}$- and $\xi$-fits. In each fit, the correlation between 
$m_\pi^2/m_q$ and $f_\pi$ for common sea quark mass is taken into account.
Between the $x$- and $\hat{x}$-fit, all of the resulting fit parameters are 
consistent. Among them, $B$ and $f$, the LECs at the leading 
order ChPT, are also consistent with the $\xi$-fit.
This indicates that the NLO formulae successfully describes the data. 

The agreement among the different expansion prescriptions is lost 
(with the deviation greater than 3$\sigma$) when we extend the fit range to
include the next lightest data point at $m_\pi\simeq$ 520~MeV.
We, therefore, conclude that for these quantities the NLO ChPT may be 
safely applied only below $\approx$ 450~MeV.

Another important observation from Figure~\ref{FIT3pts} is that only
the $\xi$-fit reasonably describes the data beyond the fitted region.
With the $x$- and $\hat{x}$-fits the curvature due to the chiral
logarithm is too strong to accommodate the heavier data points.
In fact, values of the LECs with the $x$- and $\hat{x}$-fits 
are more sensitive to the fit range than the $\xi$-fit.
This is because $f$, which is significantly smaller than $f_\pi$ of our 
data, enters in the definition of the expansion parameter.
Qualitatively, by replacing $m_q$ and $f$ by $m_\pi^2$ and $f_\pi$ 
the higher loop effects in ChPT are effectively resummed and the
convergence of the chiral expansion is improved.

We then extend the analysis to include the NNLO terms.
Since we found that only the $\xi$-fit reasonably describes the data
beyond $m_\pi\simeq$ 450~MeV, we perform the NNLO analysis using the
$\xi$-expansion in the following.
With other expansion parameters, the NNLO fits including heavier mass
points are unstable.
At the NNLO, the formulae in the $\xi$-expansion are \cite{Colangelo:2001df}
\begin{align}
  m_\pi^2/m_q & = 2B
    \Bigl[
    1 +\tfrac{1}{2}\xi\ln\xi+\tfrac{7}{8}(\xi\ln\xi)^2
  \nn\\
  &  
  +\left(
    \tfrac{c_4}{f} -\tfrac{1}{3}(\tilde{l}^{\rm\ phys}+16)
  \right)\xi^2 \ln\xi
  \Bigr]\nn\\
  &  +c_3\, \xi(1-\tfrac{9}{2}\xi\ln\xi)
  +\alpha\,\xi^2,\label{Mchiral}
  \\
  f_\pi = f &
    \left[
      1 -\xi\ln\xi +\tfrac{5}{4}(\xi\ln\xi)^2 
      +\tfrac{1}{6}(\tilde{l}^{\rm\ phys} +\tfrac{53}{2})\xi^2\ln\xi
    \right]
  \nn\\
  & 
  +c_4\,\xi(1-5\xi\ln\xi) 
 +\beta\,\xi^2.\label{Fchiral}
\end{align}
In the terms of $\xi^2\ln\xi$, the LECs at NLO appear: 
$\tilde{l}^{\rm phys} 
\equiv 7\,\bar{l}_1^{\rm\ phys} +8\,\bar{l}_2^{\rm\ phys}
  -15\ln (2\sqrt{2}\pi f_\pi^{\rm phys}/m_{\pi^+})^2$,
where $f_\pi^{\rm phys} = 130.7$ MeV.
We input the phenomenological estimate $\tilde{l}^{\rm phys} =
-32.0\pm 4.3$ to the fit. 
Since the data are not precise enough to discriminate between
$\xi^2\ln\xi$ and $\xi^2$ in the given region of $\xi$ (0.06$\sim$0.19),
fit parameters $\alpha$ and $\beta$ partially absorb the 
uncertainty in $\tilde{l}^{\rm phys}$. In fact, our final
results for the LECs is insensitive to $\tilde{l}^{\rm phys}$.

\begin{figure}
  \begin{center}
    \includegraphics[width=7.7cm,clip]{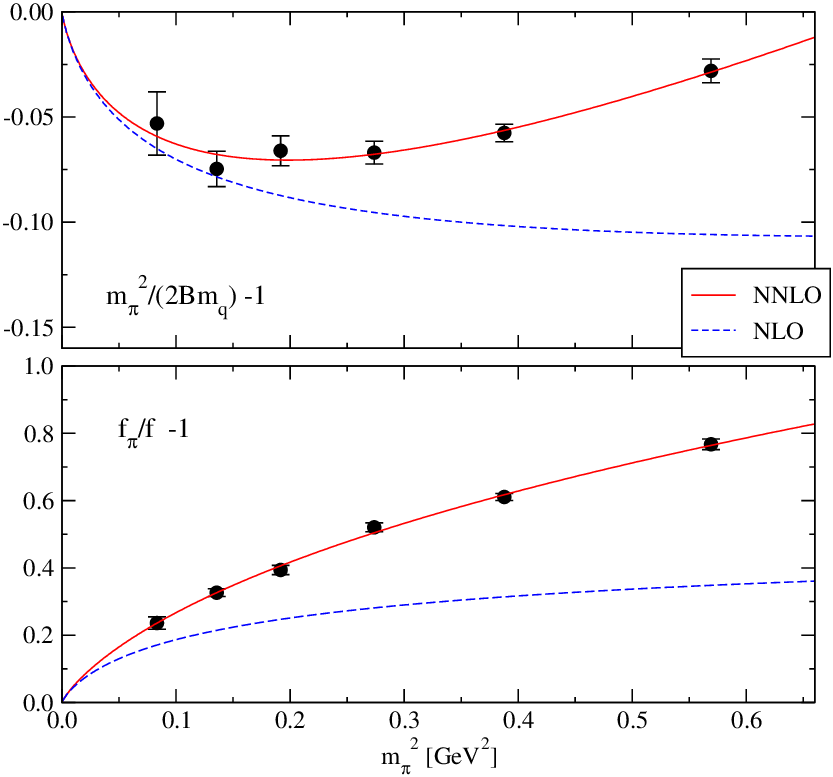}
  \end{center}
 \vspace{-0.4cm}
 \caption{
   NNLO chiral fits using all the data points 
   for $m_\pi^2/m_q$ (top) and $f_\pi$ (bottom). 
   Data are normalized by the value in the chiral limit.
   Solid curves show the NNLO fit, and the truncated expansions at NLO
   are shown by dashed curves.
   }
   \label{FIT_NNLO}
\end{figure}

In Figure~\ref{FIT_NNLO}, we show the NNLO fits using all the data
points (solid curves). 
In these plots $m_\pi^2/m_q$ and $f_\pi$ are normalized by their
values in the chiral limit.
As expected from the good convergence of the $\xi$-fit even at NLO, the NNLO
formulae nicely describe the lattice data in the whole data region.
We also draw a truncation at the NLO level (dashed curves)
but using the same fit parameters. 
The difference between the NLO truncated curves and the 
NLO fit curves to the three lightest data points
(Figure~\ref{FIT3pts}) is explained by the presence of the
terms $\xi(1-\tfrac{9}{2}\xi\ln \xi)$ and $\xi(1-5\xi\ln\xi)$ in 
(\ref{Mchiral}) and (\ref{Fchiral}), respectively.  
Since the factors $(1-\tfrac{9}{2}\xi\ln \xi)$ and $(1-5\xi\ln\xi)$ are
significantly larger than 1 in the data region, the resulting fit
parameters $c_3$ and $c_4$ in the NNLO formulae are much lower than
those of the NLO fits. 
This indicates that the determination of the NLO LECs is quite
sensitive to whether the NNLO terms are included in the analysis,
while the leading order LECs are stable.

From Figure~\ref{FIT_NNLO} we can explicitly observe the convergence
behavior of the chiral expansion.
For instance, at the kaon mass region $m_\pi\sim$ 500~MeV, the NLO
term contributes at a $-10\%$ ($+28\%$) level to $m_\pi^2/m_q$
($f_\pi$), and the correction at NNLO is about $+3\%$ ($+18\%$).
At least, the expansion is converging (NNLO is smaller than NLO) for 
both of these quantities, but quantitatively the convergence behavior
depends significantly on the quantity of interest.
For $f_\pi$ the NNLO contribution is already substantial at the kaon
mass region.

\begin{figure}
  \includegraphics[width=8.1cm,clip]{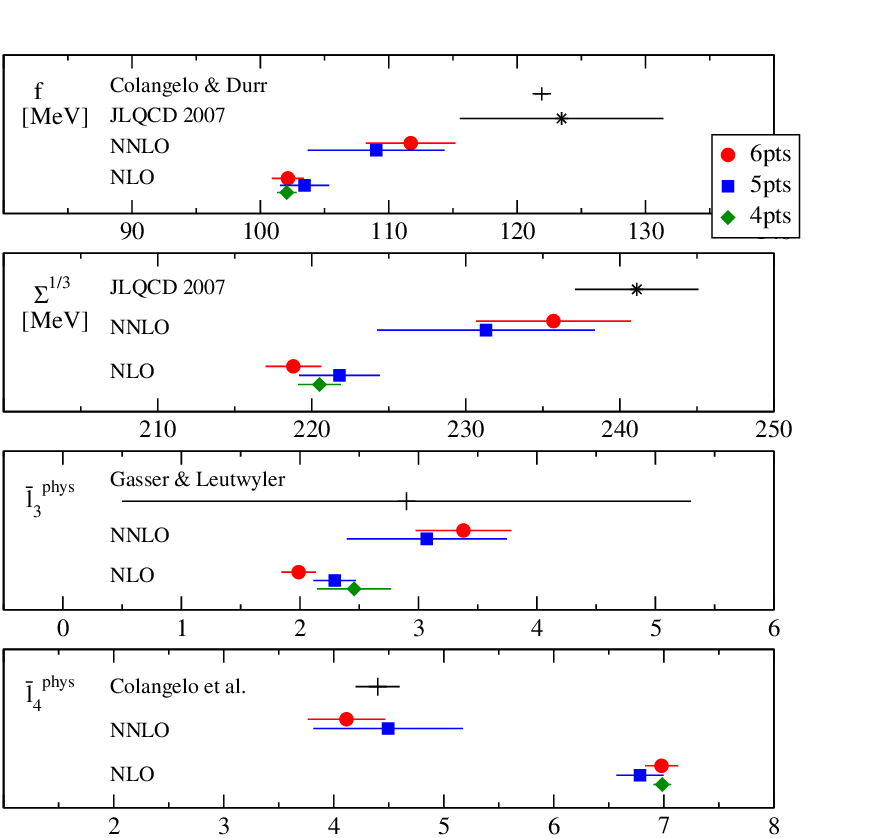}
  \caption{
 Comparison of the ChPT parameters 
 obtained from the NLO fit and NNLO fit. 
 Red circles, blue squares and green diamonds are corresponding to the 
 results of the fit using the lightest 4, 5 and all 6 data points. 
 Error bars indicate statistical error.
 Pluses and star represent reference points.}
  \label{NLOvsNNLO}
\end{figure}

From the $\xi$-fit, we extract the LECs of ChPT, {\it i.e.} 
the decay constant in the chiral limit $f$, chiral condensate
$\Sigma= Bf^2/2$, and the NLO LECs
$\bar{l}_3^{\rm phys}= -c_3/B +\ln (2\sqrt{2}\pi f/m_{\pi^+})^2$ and 
$\bar{l}_4^{\rm phys}= c_4/f +\ln (2\sqrt{2}\pi f/m_{\pi^+})^2$.
For each quantity, a comparison of the results between the NLO and the 
NNLO fits is shown in Figure~\ref{NLOvsNNLO}. 
In each panel, the results with 5 and 6 lightest data points are plotted
for the NNLO fit. 
The correlated fits give $\chi^2/$dof = 1.94 and 1.40, respectively. 
For the NLO fits, we plot results obtained with 4, 5 and 6 points to
show the stability of the fit. The $\chi^2/$dof is less than 1.94.
The results for these physical quantities are consistent within either
the NLO or the NNLO fit. On the other hand, as seen for 
$\bar{l}_4^{\rm phys}$ most prominently, there is a significant 
disagreement between NLO and NNLO. 
This is due to the large NNLO coefficients as already discussed.

We quote our final results from the NNLO fit with all data
points:
$f=111.7(3.5)(1.0)(^{+6.0}_{-0.0})$~MeV,
$\Sigma^{\ovl{\rm MS}}(\mathrm{2~GeV})=
 [235.7(5.0)(2.0)(^{+12.7}_{-\ 0.0})\mathrm{~MeV}]^3$, 
$\bar{l}_3^{\rm phys}=3.38(40)(24)(^{+31}_{-\ 0})$, and
$\bar{l}_4^{\rm phys}=4.12(35)(30)(^{+31}_{-\ 0})$.
From the value at the neutral pion mass $m_{\pi^0}=135.0$ MeV, 
we obtain the average up and down quark mass $m_{ud}$ and the pion
decay constant as 
$m_{ud}^{\ovl{\rm MS}}(\mathrm{2~GeV})
=4.452(81)(38)(^{+\ 0}_{-227})$~MeV
and $f_\pi =119.6(3.0)(1.0)(^{+6.4}_{-0.0})$~MeV.
In these results, the first error is statistical, 
where the error of the renormalization constant is included in quadrature 
for $\Sigma^{1/3}$ and $m_{ud}$.
The second error is systematic due to the truncation of the higher
order corrections, which is estimated by an order counting with a
coefficient of $\approx 5$ as appeared at NNLO.
For quantities carrying mass dimensions, the third error is from 
the ambiguity in the determination of $r_0$. 
We estimate these errors from the difference of the results with our
input $r_0=0.49$~fm and that with $0.465$~fm \cite{Aubin:2004fs}. 
The third errors for $\bar{l}_3^{\rm phys}$ and $\bar{l}_4^{\rm phys}$
reflect an ambiguity of choosing the renormalization scale of ChPT
($4\pi f$ or $4\pi f_\pi$). 
There are other possible sources of systematic errors that are not
reflected in the error budget. They include the discretization effect,
remaining finite volume effect and the effect of missing strange quark 
in the sea.

In each panel of Figure~\ref{NLOvsNNLO}, we also plot reference points
(pluses and star) for comparison.
Overall, with the NNLO fits, we find good agreement with those
reference values.
For $f$, our result is significantly lower than the two-loop result 
in two-flavor ChPT \cite{Colangelo:2003hf}, $f= 121.9\pm 0.7$~MeV, but taking
account of the scale uncertainty, which is not shown in the plot, the
agreement is more reasonable. 
For $f$ and $\Sigma^{\overline{\mathrm{MS}}}(\mathrm{2~GeV})$, 
we also plot the lattice results from our independent simulation in the 
$\epsilon$-regime \cite{Fukaya:2007pn}.
We observe a good agreement with the NNLO fits.
Comparison of the LECs 
$\bar{l}_3^{\rm phys}$ and $\bar{l}_4^{\rm phys}$ with the
phenomenological values \cite{Colangelo:2001df} also favor the NNLO
fits, especially for $\bar{l}_4^{\rm phys}$.

With the presently available computational power, the chiral
extrapolation is still necessary in the lattice QCD calculations.
The consistency test of the lattice data with ChPT as described in
this paper is crucial for reliable chiral extrapolation of any
physical quantities to be calculated on the lattice.
With a two-flavor simulation preserving exact chiral symmetry,
we demonstrate that the lattice data are well described with the use
of the resummed expansion parameter $\xi=2m_\pi^2/(4\pi f_\pi)^2$.
Extension of the analysis to the case of partially quenched QCD 
\cite{Nf2SpectFull} and to other physical quantites, such as the
pion form factor \cite{Kaneko:2007nf} is on-going.
Also, simulations with exact chiral symmetry including dynamical
strange quark are underway~\cite{Hashimoto:2007vv}.

\begin{acknowledgments}
Numerical simulations are performed on Hitachi SR11000 and
IBM System Blue Gene Solution at High Energy Accelerator Research
Organization (KEK) under a support of its Large Scale
Simulation Program (Nos.~07-16). HF was supported by Nishina Foundation. 
This work is supported in part by the Grant-in-Aid of the
Ministry of Education
(Nos.
17740171,
18034011,
18340075,
18740167,
18840045,
19540286,
19740121,
19740160,
20025010,
20039005,
20340047,
20740156)
and the National Science Council of Taiwan (Nos.~NSC96-2112-M-002-020-MY3,
NSC96-2112-M-001-017-MY3).
\end{acknowledgments}

\bibliography{SpectNf2b230ref}

\end{document}